\documentclass[prd, preprint, amsmath, amssymb]{revtex4}

\usepackage{bm}
\usepackage{graphicx}
\usepackage{amsmath}
\usepackage{latexsym}

\newcommand{\Tr}{\mathrm{Tr}}
\newcommand{\tr}{\mathrm{tr}}
\newcommand{\point}{\; .}

\begin{document}


\title{An insight on the Proof of Orientifold Planar Equivalence from the Lattice}

\author{Agostino Patella}
\affiliation{Scuola Normale Superiore, Pisa}
\affiliation{INFN, Pisa}
\email[E-mail me at: ]{patella@sns.it}

\date{04/21/2006}

\begin{abstract}
In a recent paper, Armoni, Shifman and Veneziano (ASV) gave a formal non-perturbative proof of planar equivalence between the bosonic sectors of  $SU(N)$ super Yang-Mills theory and of a gauge theory with a massless quark in the antisymmetric two-indexes representation. In the case of three colors, the latter theory is nothing but one-flavor QCD. Numerical simulations are necessary to test the validity of that  proof and to estimate the size of  $1/N$ corrections. As a first step towards numerical simulations, I will give a lattice version of the ASV  proof of orientifold planar equivalence in the strong-coupling and large-mass phase.
\end{abstract}

\maketitle


\section{Introduction}

The understanding of gauge theories in the strong coupling regime is one of the most challenging problems of theoretical particle  physics. Supersymmetric theories are unique in their kind. Actually, exact results were obtained in $ \mathcal{N} = 1 $ SYM, like the gluino condensate \cite{shifman:condensate}, the NSVZ beta function \cite{novikov:beta} and the tension of domain walls \cite{dvali:domainwalls}; moreover, quark confinement and dual Meissner effect were demonstrated in some supersymmetric theories, like in the celebrated Seiberg-Witten solution \cite{seiberg:N2}. On the other hand, we still understand little about QCD in the strong coupling regime.

In a recent work, Armoni, Shifman and Veneziano (ASV) \cite{armoni:planar,armoni:relics} argued that some non-supersymmetric gauge theories are equivalent, in some appropriate sectors, to super Yang-Mills in the large-$N$ limit. I will point out the equivalence in the large-$N$ limit of the SYM (parent theory) and the gauge theory with a massless quark in the antisymmetric two-indexes representation (daughter theory). The latter may appear somewhat strange; but, in the case of three colors, the antisymmetric representation is the same as the antifundamental one, and the daughter theory is nothing but one flavor QCD. Thus, exact results in supersymmetry can be used to estimate the corresponding quantities in one flavor QCD.

At this point, more precise statements are needed. In practice, the large-$N$ expansion of a theory is performed as a topological expansion of its graphs. The large-$N$ limit is usually identified with the resummation of all the planar graphs. This requires the interchange of the large-$N$ limit with the sum of infinite graphs, which is not always allowed. Therefore, I will refer to the resummation of all the planar graphs as the planar limit of the theory, in order to distinguish it from the correct large-$N$ limit.

The \textit{orientifold planar equivalence} is precisely the equivalence of the two theories in the planar limit. In order to obtain meaningful predictions in one flavor QCD by the orientifold planar equivalence, two conditions must be fulfilled: the planar limit must coincide with the large-$N$ limit and the $1/N$ corrections must be small.

A proof of orientifold planar equivalence on continuum is given by Armoni, Shifman, Veneziano in \cite{armoni:proof}. Even if no analytic control of the previous conditions is available, an interesting evaluation of quark condensate has been obtained \cite{armoni:condensate}. I expect future numerical simulations to give precise and meaningful informations on these topics.

This paper is a preliminary work to numerical simulations. I will discuss the orientifold planar equivalence of the following two theories on the lattice:
\begin{description}
\item[AdjQCD.] Gauge theory with one Majorana fermion of mass $m$ in the adjoint representation.
\item[AsQCD.] Gauge theory with one Dirac fermion of mass $m$ in the antisymmetric two-indexes representation.
\end{description}
It will be clear that in such gauge theories, the graphs can be easly defined by expanding the expectation values of gauge-invariant observables in the bosonic sector as a power series of $1/g^2$ and $1/m$. The orientifold planar equivalence can be proved graph by graph, in a similar way like in Armoni, Shifman, Veneziano \cite{armoni:proof}.

It is worth to repeat that this planar limit doesn't \textit{a priori} coincide with the large-$N$ limit. It is known that the pure gauge theory in two dimensions on the lattice has a phase transition in the large-$N$ limit (at finite volume, too) between a strong-coupling and a weak-coupling phase \cite{gross:2d}. In that case, the planar limit coincides with the large-$N$ limit only in the strong-coupling phase. Likewise, the orientifold planar equivalence shows the equivalence of \textbf{AdjQCD} and \textbf{AsQCD} in the large-$N$ limit only in a strong-coupling and large-mass phase, leaving opened the question if it extends in a more physical region, too.

In Section \ref{lattice}, I will review the notations for a gauge theory on the lattice. In Section \ref{hopping}, I will expand the fermionic effective action as a sum of Wilson loops by means of the hopping expansion; this is a preliminary step towards the expansion of the expectation values of the observables as a power series in the hopping parameter (essentially $1/m$). In Section \ref{graphs}, I will perform the full graphs expansion of both the considered theories. Finally in Section \ref{planar}, I will perform the planar limit and I will prove the orientifold planar equivalence in the strong-coupling and large-mass phase.


\section{Orientifold planar equivalence on the lattice\label{lattice}}

The dynamical degrees of freedom of a gauge theory on the lattice are the link variables $ U_{\hat{\mu}}(x) $, defined as the Wilson lines of the gauge field along the straight path from the site $ x $ to its nearest neighbouring site $ x+\hat{\mu} $:
\begin{equation}
U_{\hat{\mu}}(x) = \textrm{Pexp} \, i \int_0^1 A(x+as\hat{\mu}) \, ds \point
\end{equation}
In what follows, because I am not interested in the continuum limit, I will fix the lattice spacing $ a=1 $.

The action for a pure gauge theory is the Wilson action
\begin{equation} \label{wilsonaction}
S_W(\lambda, U) = - \frac{2 N^2}{\lambda} \sum_p \left( 1- \frac{1}{N} \textrm{Re} \, \tr \, U_p \right)
\end{equation}
where $ \sum_p $ is the sum over all the plaquettes (squared paths with unitary side), $ U_p $ is the ordered product of the link variables along the plaquette $ p $ and $ \lambda = g^2 N $ is the 't Hooft coupling constant.

In order to introduce quarks, it is necessary to define a discretized version of the Dirac operator. There are more than one way to do that. In this work, I will consider the Dirac operator for Wilson fermions
\begin{eqnarray}
&& D_{xy} =  \delta_{xy} - \kappa M(U)_{xy} = \nonumber \\
&& \quad = \delta_{xy} - \kappa \sum_\mu \left\{
( r-\gamma_\mu ) \bm{r}[U_{\hat{\mu}}(x)] \delta_{x+\hat{\mu}, y} +
( r+\gamma_\mu ) \bm{r}[U_{-\hat{\mu}}(x)] \delta_{x-\hat{\mu}, y} \right\} \label{diracoperator}
\end{eqnarray}
where $ \bm{r} $ is the representation of the gauge group which the quarks belong to, $ r $ is the Wilson parameter and $ \kappa = (2m+8r)^{-1} $ is the hopping parameter.

Therefore, the partition function of a gauge theory on the lattice with $ N_f $ Dirac fermions (or, equivalently, $ 2N_f $ Majorana fermions) in the representation $ \bm{r} $ is
\begin{equation}
Z(\lambda,\kappa) = \int \mathcal{D}U \; e^{-S_W(\lambda,U)} \; \det \left( \bm{1} - \kappa M_{\bm{r}}(U) \right)^{N_f} \point \label{partitionfunction}
\end{equation}

Given a gauge-invariant observable $ F $, namely a product of Wilson loops
\begin{equation} \label{observable}
F(U) = \tr \, U_{\Gamma_1} \cdots \tr \, U_{\Gamma_w}
\end{equation}
the generating functional of the expectation values of the powers of $F$ is the free energy in presence of an external source $J$ coupled with $F$
\begin{equation}
W(\lambda,\kappa,J) = - \log \int \mathcal{D}U \; e^{-S_W[U]} \; \det \left( \bm{1} - \kappa M_{\bm{r}}(U) \right)^{N_f} \; e^{iJF[U]} \point \label{freeenergy}
\end{equation}

Keeping fixed the number of colors, one can expand in power series of $\kappa$ (hopping expansion), $\lambda^{-1}$ (strong coupling expansion) and $J$
\begin{equation}
W(\lambda,\kappa,J) = \sum_{p,q,n = 0}^\infty \lambda^{-p} \kappa^q J^n W_{p,q,n}(N) \point
\end{equation}

Finally, I can formulate the claim of this paper as
\begin{equation}
\lim_{N \rightarrow \infty} W^\textbf{AdjQCD}_{p,q,n}(N) = \lim_{N \rightarrow \infty} W^\textbf{AsQCD}_{p,q,n}(N) \point
\end{equation}


\section{Hopping expansion\label{hopping}}

The fermionic contribution to the free energy (\ref{freeenergy}) can be written as a non-local effective action, by simply taking minus the logarithm of the fermionic determinant.

\begin{equation}
S_F = - \log \det \left( \bm{1} - \kappa M_{\bm{r}}(U) \right)^{N_f} = - N_f \Tr \log \left( \bm{1} - \kappa M_{\bm{r}}(U) \right)
\end{equation}

The logarithm of the Dirac operator can be defined by means of a power series in the hopping parameter.
\begin{equation}
S_F = N_f \sum_{j=1}^\infty \frac{\kappa^j}{j} \Tr M^j =
N_f \sum_{j=1}^\infty \; \sum_{\substack{x_1,\dots,x_j \\ \in \text{lattice}}} \frac{\kappa^j}{j} \tr \left( M_{x_1 x_2}M_{x_2 x_3}\cdots M_{x_j x_1} \right)
\end{equation}
If one chooses the chiral representation for the gamma matrices and $r < 1$, then the elements of $M$ are all in the circle of radius $1$ in the complex plane. Since the matrix $M$ acts on a complex vectorial space of dimension $ 4Vd_\textbf{r}$ ($V$ is the volume of the lattice, $d_\textbf{r}$ is the dimension of the fermionic representation), then
\begin{equation}
\frac{\kappa^j}{j} | \Tr M^j | \leq \frac{1}{j} \left(4Vd_\textbf{r}\kappa\right)^j \point
\end{equation}
The second member of the last disequation is the generic term of the power expansion of $ - \log \left(1-4Vd_\textbf{r}\kappa\right) $ and, since the radius of convergence of $ \log(1+x) $ is $1$, the hopping expansion (at fixed $N$) is absolutely convergent if $ 4Vd_\textbf{r}\kappa < 1 $. In this case the mass (in units of $a^{-1}$) is larger than $ 4Vd_\textbf{r} + 4 $.

Since the matrix $ M $ has non-zero elements only between nearest neighbouring sites, contributions to the sum come only from n-ples $ (x_1, x_2, \dots, x_j) $ that make closed paths linking nearest neighbouring sites on the lattice. I define $ \mathcal{C} $ as the set of such loops. Notice that loops differing only in starting point or in winding number are different elements of $ \mathcal{C} $. Moreover, chosen $ (x) $ in $ \mathcal{C} $, I define $ L(x) $ as the length of the loop and $ M(x) $ as the ordered product of $ M $ on $ (x) $. The hopping expansion can be written as:
\begin{equation}
S_F = N_f \sum_{(x) \in \mathcal{C}} \frac{\kappa^{L(x)}}{L(x)} \tr M(x) \point
\end{equation}

Now, $ M_{xy} $ is a matrix with both \mbox{spin-$ \frac{1}{2} $} and gauge indexes; but the \mbox{spin-$ \frac{1}{2} $} and the gauge parts are factorizable. Actually, the gauge part is exactly the parallel transport along the link $ (x, y) $, so I can write:
\begin{eqnarray}
S_F &=& N_f \sum_{(x) \in \mathcal{C}} c(x) \mathcal{W}(x) \\
\mbox{with } && \mathcal{W}(x) = \tr \left( U_{x_1 x_2}U_{x_2 x_3}\cdots U_{x_j x_1} \right) \nonumber \\
&& c(x) = \frac{\kappa^{L(x)}}{L(x)} \tr \left[ (r-\gamma_{x_2-x_1}) \cdots (r-\gamma_{x_1-x_L}) \right] \nonumber \\
&& \gamma_{\pm\hat{\mu}} = \pm \gamma_\mu \point \nonumber
\end{eqnarray}
This gives the expansion in Wilson loops $ \mathcal{W}(x) $ of the fermion effective action. The coefficient $ c(x) $ are real and representation indipendent.

One can write the free energy of the two theories as:
\begin{eqnarray}
W_{\textbf{AdjQCD}}[J] &=& -\log \int \mathcal{D}U \; \exp \left\{ -S_W - \frac{1}{2} \sum_{(x) \in \mathcal{C}} c(x) \mathcal{W}_{\textbf{Adj}}(x) + iJF \right\} \nonumber \\
W_{\textbf{AsQCD}}[J] &=& -\log \int \mathcal{D}U \; \exp \left\{ -S_W - \sum_{(x) \in \mathcal{C}} c(x) \mathcal{W}_{\textbf{As}}(x) + iJF \right\} \point \label{specfreeenergy}
\end{eqnarray}

They differs each other in the $\frac{1}{2}$ factor in the AdjQCD due to the Majorana fermion and in the representation of Wilson loops.

Using standard relations, Wilson loops in the adjoint and antisymmetric representations can be written in terms of Wilson loops in the fundamental and antifundamental representations:
\begin{eqnarray}
&& \frac{1}{2} \tr \, \textbf{Adj}(U) = \frac{1}{2} \left\{ \left| \tr U \right|^2 -1 \right\} \nonumber \\
&& \tr \, \textbf{As}(U) = \frac{1}{2} \left\{ \left( \tr U \right)^2 - \tr U^2 \right\} \point
\end{eqnarray}
The term $-1$ in the former relation contributes as an additive constant to the free energy, so I will ignore it. In order to prove the orientifold planar equivalence, I will show that $ \tr U^2 $ is neglectable and $ \left( \tr U \right)^2 $ can be replaced by $ \left| \tr U \right|^2 $ in the planar limit.


\section{Graphs on the lattice\label{graphs}}

In order to write the free energies in (\ref{specfreeenergy}) as a sum of graphs, I must expand the exponential and the logarithm in power series. This is the same as to perform a large $\lambda$ and small $\kappa$ expansion.
\begin{eqnarray} \label{connectedexpansion}
&& W[J] = - \log \left< e^{-S[J]} \right> = -\sum_{N = 1}^{\infty} \frac{(-1)^N}{N!} \left< S[J]^N \right>_c \\
&& \left< S^N \right>_c = \sum_{\substack{a_1 \dots a_N = 0 \\ \sum ia_i = N}}^\infty (-1)^{\sum a_i-1} \frac{ \left( \sum a_i -1 \right)! N!}{\prod \left\{ a_i! (i!)^{a_i} \right\} } \prod \left< S^i \right>^{a_i} \point \nonumber
\end{eqnarray}
Here the brackets $ \left< \cdot \right> $ stands for the integration with respect to the Haar measure in each link. $ \left<S^N\right>_c $ is a connected expectation value, and it is defined by the expression above. In the usual perturbative expansion of a continuum quantum field theory, only connected graphs contribute to the connected expectation value. On the lattice, the connected expectation values are not related to topological connection in a simple way, as I will see later.

Now, I can replace $S[J]$ with its expression and expand all the powers $S^i$. In this way, each $\left< S^i \right>$ factor in the (\ref{connectedexpansion}) gives rise to a sum of expectation values (with respect to the Haar measure in each link) of finite products of link variables. So, in order to compute each term of the free energy, I must compute integrals like:
\begin{equation}
\int U_{i_1 j_1} \cdots U_{i_m j_m} U^*_{k_1 l_1} \cdots U^*_{k_n l_n} \, dU
\end{equation}
This can be done \cite{collins:haar}, but in this way the free energy has a tremendously complex expression. Because I'm interested in the planar limit, I will keep only the leading terms in $1/N$ of the previous integral. It can be seen that at leading order only integrals with the same number of $U$ and $U^\dagger$ contribute.
\begin{equation} \label{integrals}
\int U_{i_1 j_1} \cdots U_{i_m j_m} U^*_{k_1 l_1} \cdots U^*_{k_m l_m} \, dU = \frac{1}{N^m} \sum_{\sigma \in \mathfrak{S}_m} \delta_{i_1 k_{\sigma(1)}} \dots \delta_{i_m k_{\sigma(m)}} \delta_{j_1 l_{\sigma(1)}} \dots \delta_{j_m l_{\sigma(m)}} + O\left( \frac{1}{N^{m+1}} \right)
\end{equation}
The set $\mathfrak{S}_m$ is the group of the permutations of $m$ elements. This result can be stated as: the effect at the leading order of the integration over the gauge group is to perform Wick-contractions between couples of $U$ and $U^\dagger$; each Wick-contraction give rise to a $1/N$ factor \footnote{This result can be found, for example, in \cite{collins:haar}. Probably, it was well-known from before: 't Hooft uses it in his \cite{'thooft:largeN}, though he doesn't cite his source.}.

Therefore, at this stage of approximation, the Haar measure looks very like a gaussian measure. So, one can use all the machinery developped in the usual perturbation expansion in order to classify the term of the free energy (\ref{connectedexpansion}).

Let's consider one of the terms obtained from the expansion of all the products in $\left< S^i \right>$. It is the expectation value with respect the Haar measure in each link of a number of plaquettes coming from $ S_W $, a number of Wilson loops coming from the fermionic determinant, a number of Wilson loops coming from the observable $ F $ and a number of sources $J$ (times an overall factor). The expectation value is computed by taking all the possible Wick-contraction of $U$'s and $U^\dagger$'s at the same link and summing over all the ways of doing that. Each term in this sum is a graph.

\begin{figure}
\includegraphics[width=.4\textwidth]{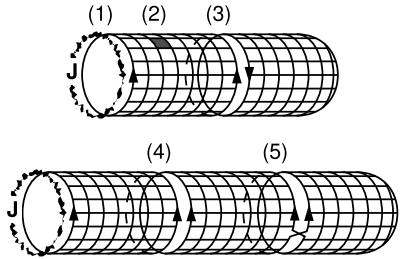}
\caption{\label{graph} Above, a graph of the free energy for AdjQCD. Below, a graph of the free energy for AsQCD. (1) External source term $JF$. (2) Plaquette. (3) $ \left( \tr U \right)^2 $ from a Wilson loop in the antisymmetric representation (from the fermionic determinant). (4) $ \left| \tr U \right|^2 $ from a Wilson loop in the adjoint representation (from the fermionic determinant). (5) $ \tr U^2 $ from a Wilson loop in the antisymmetric representation (from the fermionic determinant).}
\end{figure}

 The diagrammatic representation of a graph is obtained with the following rules (see figure \ref{graph}):
\begin{enumerate}
\item A link variable $U_{xy}$ is drawn like an arrow from the point $x$ to $y$.
\item A plaquette is drawn like an oriented, squared surface with unitary side.
\item A Wilson loop in the fundamental representation is a closed, oriented path, connecting neighbouring points on the lattice. The same Wilson loop in the antifundamental representation is the opposite-oriented path. I will consider the Wilson loops from the fermionic determinant and the observable $F$ like empty paths, unlike the plaquettes.
\item A Wilson loop in the adjoint representation is the product of a Wilson loop in the fundamental representation and in the antifundamental one; it is drawn like two closed, superimposed, opposite-oriented paths.
\item A Wilson loop in the antisymmetric representation gives rise to two terms. The first one is the product of two Wilson loops in the fundamental (or in the antifundamental) representation; it is drawn like two closed, superimposed, parallel paths. The other one is a Wilson loop in the fundamental (or in the antifundamental) representation, winding twice along the path.
\item The source $J$ is drawn like a wavy line, coupled to one or more Wilson loops.
\item If two link variables are Wick-contracted, I consider the Wilson loops or the plaquettes, which the link variables belong to, as sewn together in that link.
\end{enumerate}

In conclusion, a graph is an intricate and possibly disconnected surface tiled by plaquettes, bounded by Wilson loops and cut by couples of superimposed Wilson loops.

Let's consider now a connected expectation value in the free energy (\ref{connectedexpansion}). \textit{A priori}, it is polynomial in graphs. In the usual perturbation expansion of quantum field theory, this is not the case. In fact the expansion of the logarithm gives rise to the right signs and the right combinatorial factors, in such a way that the free energy can be written as the sum of all the connected graphs. Such graphs are those that cannot be written as a product of two other graphs (eventually times a combinatorial factor). The name "connected" comes from the fact that they are represented as graphs connected by arcs.

Since this result is based only on the fact that the integration measure is a gaussian one, it must be valid for the free energies in (\ref{specfreeenergy}) in the planar limit, with a difference. In this case, a graph which cannot be written as a product of two subgraphs (eventually times a combinatorial factor) can be represented by a disconnected surface.

At this stage, I have developped all the tools, necessary to compute the full planar limit.


\section{Planar limit\label{planar}}

Now, I want to compute the full planar limit. First of all, it must be noticed that the following approximation holds:
\begin{equation}
\tr \, \textbf{As}(U) \simeq \frac{1}{2} \left( \tr U \right)^2 \point
\end{equation}
In fact, for every graph containing the term $ \tr U^2 $, a graph in which the term $ \tr U^2 $ is replaced by $ \left( \tr U \right)^2 $ does exist. The former is suppressed by a factor $1/N$, because it contains one trace less than the latter.

As in the usual 't Hooft expansion \cite{'thooft:largeN}, graphs are proportional to $ N^\chi $, where $ \chi $ is the Euler characteristic of the surface:
\begin{equation}
\chi = 2C - 2H - B \point
\end{equation}
where $C$ is the number of connected components, $H$ is the number of handles and $B$ is the number of boundaries. In the planar limit, only graphs with the highest $\chi$ contribute.

\begin{figure}
\includegraphics[width=.4\textwidth]{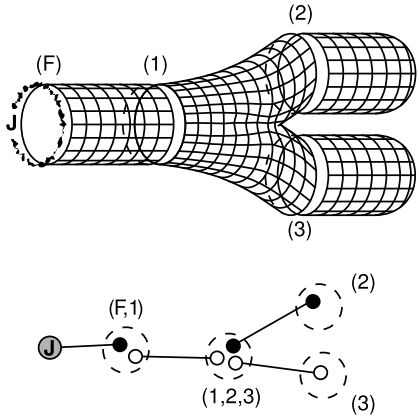}
\caption{\label{asv} A graph of the free energy and, below, its ASV representation. The clusters are represented as dashed circles.}
\end{figure}

At this point, it is useful to associate a (one-dimensional) graph like in Armoni-Shifman-Veneziano \cite{armoni:proof} to each bidimensional graph with the following rules (see figure \ref{asv}):
\begin{enumerate}
\item Each connected component corresponds to a cluster in ASV-graphs.
\item Each connected component is bounded by oriented Wilson loops, either in the fundamental or in the antifundamental representation. Each cluster contains a white circle for each Wilson loop in the fundamental representation and a black circle for each Wilson loop in the antifundamental one. So, a cluster is given by the expectation value of a product of Wilson loops and plaquettes.
\item Each couple of Wilson loops from the fermionic determinant corresponds to a line in the ASV-graphs. So, such a line contributes to the graph with the proper coefficient $ c(x) $ of the hopping expansion.
\item Each external source corresponds to a gray circle in ASV-graph. So, each gray point contributes with a $ iJ $ factor.
\item Each external source is coupled to one or more Wilson loops. In the ASV-graphs, the gray circle is connected with a line to the corresponding Wilson loops.
\end{enumerate}
In summary, a graph is given by a product of clusters, coefficients $ c(x) $, external sources and a combinatorial factor.

With this representation in hands, the graphs contributing to the free energy are easly identified. In fact, an ASV-graph cannot be written as a product of two subgraphs if and only if it is connected by arcs. Having chosen a graph, let $n$ be the number of external sources and let $w$ be the number of Wilson loops in the observable $F$, as defined in (\ref{observable}). Since the corresponding ASV-graph has $C+n$ vertices and $(B+wn)/2$ lines, its number of loops $l$ is:
\begin{equation}
l = \frac{B+wn}{2}-(C+n)+1
\end{equation}
and by using the definition of $\chi$:
\begin{equation}
\chi = -2l-2H+(w-2)n+2 \point
\end{equation}

To obtain the highest value possible for the Euler characteristic, being $w$ and $n$ fixed, One has to choose $ H = 0 $ and $ l = 0 $. So planar graphs have no handles and they are tree graphs in ASV-representation. I will use the latter property to prove the planar equivalence.

In ASV-graphs for AdjQCD, lines link points of opposite colors, instead in ASV-graphs for AsQCD lines link points of the same color. I want to show that, for each planar ASV-graph in AdjQCD, a planar ASV-graph in AsQCD of the same value does exist. Let us first see that in an example.

Consider the graph in figure \ref{asv}. Apart from the combinatorial factor, its value is:
\begin{equation}
\mathcal{G}_{\textbf{As}} = iJ c_1 c_2 c_3 \left< F \, \mathcal{W}_{\textbf{As}}^{(1)} \, \mathcal{W}_{\textbf{As}}^{(2)\dagger} \, \mathcal{W}_{\textbf{As}}^{(3)} \, \mathcal{P}^{(F,1)} \, \mathcal{P}^{(1,2,3)} \, \mathcal{P}^{(2)} \, \mathcal{P}^{(3)} \right>_c
\end{equation}
where $ \mathcal{W}_{\textbf{As}} $ is a Wilson loop in the antisymmetric representation from the fermionic determinant, $c$ is a coefficient of the hopping expansion and $\mathcal{P} $ is a product of plaquettes. In the planar limit, I obtain:
\begin{eqnarray}
\mathcal{G}_{\textbf{As}} \simeq && \frac{1}{8} iJ c_1 c_2 c_3 \left< \tr U^{(F)} \, \mathcal{P}^{(F,1)} \, \tr U^{(1)} \right> \; \nonumber\\
&& \left< \tr U^{(1)} \, \tr U^{(2)\dagger} \, \tr U^{(3)} \, \mathcal{P}^{(1,2,3)} \right> \; \left< \tr U^{(2)} \, \mathcal{P}^{(2)} \right> \; \left< \tr U^{(3)} \, \mathcal{P}^{(3)} \right>  \point
\end{eqnarray}
Because the integration measure is invariant under the transformation $ U \rightarrow U^\dagger $, the following equality holds:
\begin{eqnarray}
&& \left< \tr U^{(1)} \, \tr U^{(2)\dagger} \, \tr U^{(3)} \, \mathcal{P}^{(1,2,3)} \right> = \left< \tr U^{(1)\dagger} \, \tr U^{(2)} \, \tr U^{(3)\dagger} \, \mathcal{P}^{(1,2,3)\dagger} \right> \\
&& \mathcal{G}_{\textbf{As}} \simeq \frac{1}{8} iJ c_1 c_2 c_3 \left< \tr U^{(F)} \, \mathcal{P}^{(F,1)} \, \tr U^{(1)} \right> \nonumber \\
&& \qquad \left< \tr U^{(1)\dagger} \, \tr U^{(2)} \, \tr U^{(3)\dagger} \, \mathcal{P}^{(1,2,3)\dagger} \right> \; \left< \tr U^{(2)} \, \mathcal{P}^{(2)} \right> \; \left< \tr U^{(3)} \, \mathcal{P}^{(3)} \right>
\end{eqnarray}
and this is nothing but the planar limit of the following graph in AdjQCD:
\begin{equation}
\mathcal{G}_{\textbf{Adj}} = \frac{1}{8} iJ c_1 c_2 c_3 \left< F \, \mathcal{W}_{\textbf{Adj}}^{(1)} \, \mathcal{W}_{\textbf{Adj}}^{(2)} \, \mathcal{W}_{\textbf{Adj}}^{(3)} \, \mathcal{P}^{(F,1)} \, \mathcal{P}^{(1,2,3)\dagger} \, \mathcal{P}^{(2)} \, \mathcal{P}^{(3)} \right>_c \point
\end{equation}

\begin{figure}
\includegraphics[width=.4\textwidth]{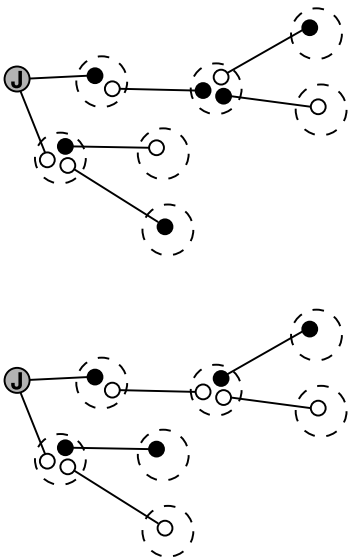}
\caption{\label{fig2} Two corresponding graphs, respectively in AdjQCD and AsQCD. It is possible to trasform one in the other, by inverting the colors of the clusters which are 2 levels far from the external source.}
\end{figure}

This result has a general validity. Since the integration measure is invariant under the substitution $ U \rightarrow U^\dagger $, one can invert all the colors in a cluster, without changing the value of the graph. Because leading ASV-graphs have no loops, one can independently choose which clusters to invert. Inverting the colors of some clusters is the same as interchanging $ \mathcal{W}_{\textbf{As}} \leftrightarrow \frac{1}{2} \mathcal{W}_{\textbf{Adj}} $. In this way, one can change the representation of the fermion and interchange Dirac with Majorana fermion. No change in the coefficients $ c(x) $ is needed, because those coefficients are representation-independent. Thus, one can associate to each ASV-graph for AdjQCD, an ASV-graph for AsQCD of the same value, properly inverting the colors of clusters (fig. \ref{fig2}).

In conclusion, the planar equivalence comes from the graph by graph equality of the free energy of the two theories. It is interesting to notice that, if $ F $ is a product of Wilson loops, the property $ l = 0 $ implies that the expectation value of $ F $ factorizes in the product of expectation values of Wilson loops.

\section{Conclusions}

In summary, I proved the large-$N$ equivalence of the lattice gauge theory with a massive Majorana fermion in the adjoint representation and of the lattice gauge theory with a massive Dirac fermion in the antisymmetric representation in the strong-coupling and large-mass phase. The proof is independent of the lattice extension. Although the proof was obtained using Wilson fermions, the hopping expansion can be performed also in the case of staggered fermions. Future numerical simulations will allow us to understand whether the orientifold planar equivalence can be extended beyond the strong-coupling and large-mass phase, how large are the $ 1/N $ corrections, and how they depend on the fermion mass, the lattice extension and spacing.

I want to notice that the same results could be obtained by using the loop-equations formalism, like in Kovtun, Unsal and Yaffe \cite{yaffe:orbifold} or Lovelace \cite{lovelace:master}. I chose the approach of the strong-coupling and hopping expansions, because it is the exact translation on the lattice of the proof of Armoni, Shifman and Veneziano \cite{armoni:proof}.

At the moment, I want to point out some hints for numerical simulations:
\begin{enumerate}
\item Due to the factorization property of Wilson loop, the simulation of the vev of one and two Wilson loops is almost enough to obtain considerable informations about orientifold planar equivalence.
\item The proof of equivalence works in the same way, independently of the lattice size. As a first step, simulations on small lattices are enough to test the equivalence and allow to reach larger value of $ N $ at the same time.
\item More careful (and much more time-expensive) simulations on larger lattices could be useful to investigate how does the orientifold planar equivalence work in the continuum limit.
\end{enumerate}


\begin{acknowledgments}

I wish to acknowledge Adriano Di Giacomo who has supported me during my work, Gabriele Veneziano for very useful discussions and Gabriele Vajente who helped me in the proof-correction.

\end{acknowledgments}



\begin{thebibliography}{13}
\expandafter\ifx\csname natexlab\endcsname\relax\def\natexlab#1{#1}\fi
\expandafter\ifx\csname bibnamefont\endcsname\relax
  \def\bibnamefont#1{#1}\fi
\expandafter\ifx\csname bibfnamefont\endcsname\relax
  \def\bibfnamefont#1{#1}\fi
\expandafter\ifx\csname citenamefont\endcsname\relax
  \def\citenamefont#1{#1}\fi
\expandafter\ifx\csname url\endcsname\relax
  \def\url#1{\texttt{#1}}\fi
\expandafter\ifx\csname urlprefix\endcsname\relax\def\urlprefix{URL }\fi
\providecommand{\bibinfo}[2]{#2}
\providecommand{\eprint}[2][]{\url{#2}}

\bibitem[{\citenamefont{Shifman and Vainshtein}(1988)}]{shifman:condensate}
\bibinfo{author}{\bibfnamefont{M.~A.} \bibnamefont{Shifman}} \bibnamefont{and}
  \bibinfo{author}{\bibfnamefont{A.~I.} \bibnamefont{Vainshtein}},
  \bibinfo{journal}{Nucl.\ Phys.} \textbf{\bibinfo{volume}{B296}},
  \bibinfo{pages}{445} (\bibinfo{year}{1988}).

\bibitem[{\citenamefont{Novikov et~al.}(1986)\citenamefont{Novikov, Shifman,
  Vainshtein, and Zakharov}}]{novikov:beta}
\bibinfo{author}{\bibfnamefont{V.~A.} \bibnamefont{Novikov}},
  \bibinfo{author}{\bibfnamefont{M.~A.} \bibnamefont{Shifman}},
  \bibinfo{author}{\bibfnamefont{A.~I.} \bibnamefont{Vainshtein}},
  \bibnamefont{and} \bibinfo{author}{\bibfnamefont{V.~I.}
  \bibnamefont{Zakharov}}, \bibinfo{journal}{Phys.\ Lett.}
  \textbf{\bibinfo{volume}{B166}}, \bibinfo{pages}{329} (\bibinfo{year}{1986}).

\bibitem[{\citenamefont{Dvali and Shifman}(1997)}]{dvali:domainwalls}
\bibinfo{author}{\bibfnamefont{G.~R.} \bibnamefont{Dvali}} \bibnamefont{and}
  \bibinfo{author}{\bibfnamefont{M.~A.} \bibnamefont{Shifman}},
  \bibinfo{journal}{Phys.\ Lett.} \textbf{\bibinfo{volume}{B396}},
  \bibinfo{pages}{64} (\bibinfo{year}{1997}), \eprint{hep-th/9612128}.

\bibitem[{\citenamefont{Seiberg and Witten}(1994)}]{seiberg:N2}
\bibinfo{author}{\bibfnamefont{N.}~\bibnamefont{Seiberg}} \bibnamefont{and}
  \bibinfo{author}{\bibfnamefont{E.}~\bibnamefont{Witten}},
  \bibinfo{journal}{Nucl.\ Phys.} \textbf{\bibinfo{volume}{B426}},
  \bibinfo{pages}{19} (\bibinfo{year}{1994}), \eprint{hep-th/9407087}.

\bibitem[{\citenamefont{Armoni et~al.}(2003{\natexlab{a}})\citenamefont{Armoni,
  Shifman, and Veneziano}}]{armoni:planar}
\bibinfo{author}{\bibfnamefont{A.}~\bibnamefont{Armoni}},
  \bibinfo{author}{\bibfnamefont{M.}~\bibnamefont{Shifman}}, \bibnamefont{and}
  \bibinfo{author}{\bibfnamefont{G.}~\bibnamefont{Veneziano}},
  \bibinfo{journal}{Nucl.\ Phys.} \textbf{\bibinfo{volume}{B667}},
  \bibinfo{pages}{170} (\bibinfo{year}{2003}{\natexlab{a}}),
  \eprint{hep-th/0302163}.

\bibitem[{\citenamefont{Armoni et~al.}(2003{\natexlab{b}})\citenamefont{Armoni,
  Shifman, and Veneziano}}]{armoni:relics}
\bibinfo{author}{\bibfnamefont{A.}~\bibnamefont{Armoni}},
  \bibinfo{author}{\bibfnamefont{M.}~\bibnamefont{Shifman}}, \bibnamefont{and}
  \bibinfo{author}{\bibfnamefont{G.}~\bibnamefont{Veneziano}},
  \bibinfo{journal}{Phys.\ Rev.\ Lett.} \textbf{\bibinfo{volume}{91}},
  \bibinfo{pages}{191601} (\bibinfo{year}{2003}{\natexlab{b}}),
  \eprint{hep-th/0307097}.

\bibitem[{\citenamefont{Armoni et~al.}(2005)\citenamefont{Armoni, Shifman, and
  Veneziano}}]{armoni:proof}
\bibinfo{author}{\bibfnamefont{A.}~\bibnamefont{Armoni}},
  \bibinfo{author}{\bibfnamefont{M.}~\bibnamefont{Shifman}}, \bibnamefont{and}
  \bibinfo{author}{\bibfnamefont{G.}~\bibnamefont{Veneziano}},
  \bibinfo{journal}{Phys.\ Rev.} \textbf{\bibinfo{volume}{D71}},
  \bibinfo{pages}{045015} (\bibinfo{year}{2005}), \eprint{hep-th/0412203}.

\bibitem[{\citenamefont{Armoni et~al.}(2004)\citenamefont{Armoni, Shifman, and
  Veneziano}}]{armoni:condensate}
\bibinfo{author}{\bibfnamefont{A.}~\bibnamefont{Armoni}},
  \bibinfo{author}{\bibfnamefont{M.}~\bibnamefont{Shifman}}, \bibnamefont{and}
  \bibinfo{author}{\bibfnamefont{G.}~\bibnamefont{Veneziano}},
  \bibinfo{journal}{Phys.\ Lett.} \textbf{\bibinfo{volume}{B579}},
  \bibinfo{pages}{384} (\bibinfo{year}{2004}), \eprint{hep-th/0309013}.

\bibitem[{\citenamefont{Gross and Witten}(1980)}]{gross:2d}
\bibinfo{author}{\bibfnamefont{D.~J.} \bibnamefont{Gross}} \bibnamefont{and}
  \bibinfo{author}{\bibfnamefont{E.}~\bibnamefont{Witten}},
  \bibinfo{journal}{Phys. Rev.} \textbf{\bibinfo{volume}{D21}},
  \bibinfo{pages}{446} (\bibinfo{year}{1980}).

\bibitem[{\citenamefont{Collins and Sniady}(2004)}]{collins:haar}
\bibinfo{author}{\bibfnamefont{B.}~\bibnamefont{Collins}} \bibnamefont{and}
  \bibinfo{author}{\bibfnamefont{P.}~\bibnamefont{Sniady}}
  (\bibinfo{year}{2004}), \eprint{math-ph/0402073}.

\bibitem[{\citenamefont{'t~Hooft}(2002)}]{'thooft:largeN}
\bibinfo{author}{\bibfnamefont{G.}~\bibnamefont{'t~Hooft}}
  (\bibinfo{year}{2002}), \eprint{hep-th/0204069}.

\bibitem[{\citenamefont{Kovtun et~al.}(2003)\citenamefont{Kovtun, Unsal, and
  Yaffe}}]{yaffe:orbifold}
\bibinfo{author}{\bibfnamefont{P.}~\bibnamefont{Kovtun}},
  \bibinfo{author}{\bibfnamefont{M.}~\bibnamefont{Unsal}}, \bibnamefont{and}
  \bibinfo{author}{\bibfnamefont{L.~G.} \bibnamefont{Yaffe}},
  \bibinfo{journal}{JHEP} \textbf{\bibinfo{volume}{12}}, \bibinfo{pages}{034}
  (\bibinfo{year}{2003}), \eprint{hep-th/0311098}.

\bibitem[{\citenamefont{Lovelace}(1982)}]{lovelace:master}
\bibinfo{author}{\bibfnamefont{C.}~\bibnamefont{Lovelace}},
  \bibinfo{journal}{Nucl. Phys.} \textbf{\bibinfo{volume}{B197}},
  \bibinfo{pages}{76} (\bibinfo{year}{1982}).

\end{thebibliography}
\end{document}